\documentclass[journal]{IEEEtran}
\IEEEoverridecommandlockouts
\usepackage{cite}
\usepackage{balance}
\usepackage{amsmath,amssymb,amsfonts}
\usepackage{algorithmic}
\usepackage{graphicx}
\usepackage{textcomp}
\usepackage{xcolor}
\usepackage{booktabs}
\usepackage{subfigure}
\usepackage{multirow}
\usepackage{xcolor}
\usepackage{soul}
\sethlcolor{pink}
\usepackage{tablefootnote}
\usepackage[T1]{fontenc}
\usepackage[utf8]{inputenc}
\usepackage{polski}

\DeclareMathOperator*{\concat}{\text{\Large $\parallel$}}

\def\BibTeX{{\rm B\kern-.05em{\sc i\kern-.025em b}\kern-.08em
    T\kern-.1667em\lower.7ex\hbox{E}\kern-.125emX}}
    
\begin{document}
\markboth{Journal of Biomedical and Health Informatics}%
{Shell \MakeLowercase{\textit{et al.}}: Bare Demo of IEEEtran.cls for IEEE Journals}

\title{FENet: A Frequency Extraction Network for Obstructive Sleep Apnea Detection
}

\author{Guanhua~Ye,
        Hongzhi~Yin,
        Tong~Chen,
        Hongxu~Chen,
        Lizhen~Cui,
        and~Xiangliang~Zhang
        
\thanks{
G. Ye, H. Yin and T. Chen are with the School of Information Technology \& Electric Engineering, The University of Queensland, Australia. E-mail: g.ye@uq.net.au, h.yin1@uq.edu.au, tong.chen@uq.edu.au. }
\thanks{H. Chen is with the School of Computer Science, University of Technology Sydney, Australia. E-mail: hongxu.chen@uts.edu.au.}%
\thanks{L. Cui is with the School of Software, Shandong University,  China. E-mail: clz@sdu.edu.cn.}%
\thanks{X. Zhang is with the Division of Computer
Electrical and Mathematical Sciences \& Engineering, King Abdullah University of Science and Technology, Saudi Arabia. E-mail: xiangliang.zhang@kaust.edu.sa.}%
\thanks{H. Yin is the corresponding author.}
}%

\maketitle

\begin{abstract}
Obstructive Sleep Apnea (OSA) is a highly prevalent but inconspicuous disease that seriously jeopardizes the health of human beings. Polysomnography (PSG), the gold standard of detecting OSA, requires multiple specialized sensors for signal collection, hence patients have to physically visit hospitals and bear the costly treatment for a single detection. Recently, many single-sensor alternatives have been proposed to improve the cost efficiency and convenience. Among these methods, solutions based on RR-interval (i.e., the interval between two consecutive pulses) signals reach a satisfactory balance among comfort, portability and detection accuracy. 

In this paper, we advance RR-interval based OSA detection by considering its real-world practicality from energy perspectives. As photoplethysmogram (PPG) pulse sensors are commonly equipped on smart wrist-worn wearable devices (e.g., smart watches and wristbands), the energy efficiency of the detection model is crucial to fully support an overnight observation on patients. This creates challenges as the PPG sensors are unable to keep collecting continuous signals due to the limited battery capacity on smart wrist-worn devices. Therefore, we propose a novel Frequency Extraction Network (FENet), which can extract features from different frequency bands of the input RR-interval signals and generate continuous detection results with downsampled, discontinuous RR-interval signals. With the help of the one-to-multiple structure, FENet requires only one-third of the operation time of the PPG sensor, thus sharply cutting down the energy consumption and enabling overnight diagnosis. Experimental results on real OSA datasets reveal the state-of-the-art performance of FENet.
\end{abstract}


\IEEEpeerreviewmaketitle

\section{Introduction} \label{sec:intro}
\IEEEPARstart{O}{bstructive} Sleep Apnea (OSA) is the most common disease in sleep-related breathing disorders \cite{b26}. One apnea event is defined as ``a decrease of airflow by more than 90\% from baseline over a period of more than 10 second'' \cite{b4}. The prevalence of OSA in the general adult population ranges from 6\% to 17\% \cite{b12}. Studies show that OSA can cause hyper-somnolence and fatigue, heart failure, stroke \cite{b1}, pulmonary hypertension \cite{b2} and even erectile dysfunction \cite{b3}. Moreover, OSA patients do not usually develop noticeable symptoms. the only observable symptom of OSA is simply ordinary snoring, which hinders a patient from self-diagnosis of this destructive disease. Therefore, a large number of OSA patients fail to receive appropriate treatment in time. The gold standard for diagnosing OSA is the polysomnography (PSG), which requires patients to access a dedicated sleep laboratory, where well-trained medical experts will attach multiple sensors to the patient and collect various signals while the patient is sleeping. Then, the medical experts will finalize the diagnosis based on the collected data. However, PSG-based diagnosis faces a high cost between \$1,000 and \$2,000 for just an overnight test, and the unfamiliar environment and constant connection to sensors during sleep further make it an unpleasant experience for patients\footnote{www.verywellhealth.com/what-to-expect-in-a-sleep-study-3015121}. As a result, PSG-based diagnosis is inconvenient and expensive, rendering large-scale and regular testing for this common disease infeasible.

To this end, it is highly desirable to develop a comfortable, portable and cost-effective approach to detect OSA in the daily environment. For patients, methods needing fewer sensors than PSG are naturally more suited to daily lives. Recently, with the advances in sensor technologies, single-sensor solutions based on physiological signals have been proposed. Among the physiological signals used for OSA detection, respiration-related signals are one of the most widely adopted options \cite{b11}. The most direct way to detect respiration is to measure the airflow. \cite{b5} compared the detection performance of oronasal thermal airflow (TA) sensor, nasal pressure (NP) transducer, respiratory inductance plethysmography (RIP) and tracheal sound (TS) sensor. The output of all four sensors directly reflects the breathing frequency of patients, thus becoming a strong indicator of OSA disease. Meanwhile, the collection of high-quality air flow data heavily relies on the precision of sensors. Unfortunately, studies \cite{b6, b7} have pointed out that only 60\% of such signals collected in overnight monitoring are valid for both children and adults. A common cause is that, the patients tend to subconsciously shift or remove the sensors they wear due to the foreign body sensation.  \cite{b10} proposed a method focusing on combinatorial audio features of the breathing process like non-respiratory rates and average ventilation.  Sound-based detection is intuitively the most convenient one, but its limitation in noise resistance makes it unsuitable for daily scenarios like bed-sharing.

Though existing single-sensor methods with respiratory signals fail to offer ideal real-life practicality, there is another choice, namely the electrocardiogram (ECG). Thanks to the open-source PhysioNet apnea-ECG database \cite{b11}, ECG appears to be the most popular data source for OSA detection task since 2003 \cite{b12}.  Respiration-based methods can easily establish explicit and intuitive associations between respiratory data and respiratory diseases (e.g., OSA), however, this is not the case with ECG data as its relationship with respiratory diseases can only be implicitly captured. Therefore, the design of an effective prediction algorithm is the key to successful OSA detection. ECG-based OSA detection is mainly performed in two ways. The first exploits full information of ECG recordings with techniques like wavelet transform \cite{b13, b14, b15} and variational mode decomposition \cite{b16}. In contrast, the other type of methods only need a basic yet important attribute within ECG recordings, i.e., the RR-interval. RR-interval is the time gap between two R-waves, which is identical to the interval between two heart beats. While obtaining the full ECG data during sleep is non-trivial (e.g., multiple electrodes must keep attached to a patient's chest), one major advantage of RR-interval signals is that they can be easily collected via photoplethysmogram (PPG) pulse sensors. Nowadays, PPG sensors are commonly equipped on wrist-worn wearable devices like smart watches and the much cheaper smart wristbands\footnote{https://bit.ly/smart-wrist-band}, which means we do not have to develop any new hardware system from the ground up. In this case, we avoid persuading potential OSA patients to buy OSA diagnosis devices as OSA symptoms are hardly noticeable for themselves, but simply add an extra functionality to their smart wrist-worn devices for day-to-day use.

To detect OSA from RR-interval signals, various machine learning methods are devised, including support vector machines (SVM) \cite{b15}, k-nearest neighbors (KNN) algorithm  \cite{b17}, linear discriminant analysis (LDA) \cite{b18}, hidden Markov model (HMM) \cite{b19}. With the immense advances in deep learning \cite{deeplearning}, methods based on deep neural networks \cite{b13} and recurrent neural networks (RNNs) \cite{b20} are proven effective in this task. More recently, \cite{b21} proposed a model combining convolutional neural networks (CNNs) with long- and short-term memory (LSTM) networks achieved a considerable performance, which indicates RR-interval can be a key feature for OSA detection. As a result, collecting and analyzing RR-interval signals for OSA detection is an ideal solution that offers comfort, portability, and low cost at the same time. Considering the prevalence of the aforementioned smart wearable devices, the cost of detecting OSA can be dramatically reduced. Meanwhile, the portability of wearable devices facilitates people tracking their OSA status in a real-time manner. 

Nevertheless, on top of the high demand on detection accuracy, the migration towards wrist-worn devices comes along with a new challenge. As data processing tasks can be assigned to connected smartphones or the cloud, wearable devices' major task is simply gathering RR-interval data. However, the battery capacity in smart wearable devices is commonly small (e.g., around 100 mAH in most smart wristbands), while PPG sensors are driven by an electric current of up to 30 mA (see Section \ref{sec:energy}). So in the worst case, pulse sensors can only work continuously for 3.3 hours, rendering it difficult for overnight monitoring. In this regard, the energy efficiency will be the defining quality of whether such OSA detection solution is realistic.
 Existing methods rarely take energy efficiency into consideration as they simply assume the availability of a sequence of continuous, long-range RR-interval records. In light of this, this paper proposes a detection model, namely the Frequency Extraction Network (FENet) that can produce continuous diagnostic results on discontinuous signals. This design 
enables the PPG sensor to work with an intermittent duty cycle, thus easing the pressure on the battery of wearable devices and maximizing the monitoring duration. To the best of our knowledge, this is the first work addressing the real-life practicality of OSA detection on wrist-worn devices by reducing the duty cycle of sensors. The major contributions of this paper are:

\begin{itemize}
\item We identify the unique challenges of OSA detection on smart wrist-worn devices, and design an energy-efficient machine learning solution based on RR-interval signals.
\item We propose a novel model FENet for OSA detection. With an innovative multi-frequency dilated convolutional neural network, FENet learns implicit representations of discontinuous signals, and then generates apnea labels for both the current timestamp and the adjacent ones with missing input. Thus, FENet allows for accurate OSA detection with limited data availability.
\item Extensive experiments on real-world OSA datasets show the superior accuracy of FENet. Further studies on energy efficiency suggest that FENet effectively reduces the energy consumption and is more friendly to the limited battery capacity on smart wrist-worn devices. 
\end{itemize}

\section{Related Work}
In RR-interval-based OSA detection, there have been attempts to decrease the energy consumption of wearable devices. \cite{surrel2016low} used the frequency spectrum of the RR-intervals to cut down the energy cost of feature extraction, which proves identifiable relations between frequency-based patterns and the OSA disease. \cite{surrel2018online} performed sleep apnea assessment in time domain and applied an efficient outlier removal technique to reduce the time complexity of OSA detection, enabling energy saving. However, the handcrafted features are highly dependent on domain expertise and are hard to generalize to other large-scale datasets.

End-to-end deep learning models are applied to bypass the need for manual feature selection. \cite{baseline_CNN} proposed a method that recognizes apnea events via a multi-layer CNN model, which can capture high-level semantic features automatically. RNN models have been intensively studied and perform well in sequential representations \cite{chen2020sequence, chen2018tada, KDD19Streaming, WWW20Next}. \cite{baseline_LSTM} used bidirectional RNNs to learn a hidden state to represent RR-interval series and then classify input with this feature. Liang \textit{et al}. improved the overall performance by combining CNNs and RNNs, where the RR-interval signals have to pass through two convolutional layers and one LSTM layer \cite{b21}. Regrettably, these works mainly focus on how to achieve a higher accuracy without considering the differences in operational conditions between wearable devices and sleep laboratories. 

As pointed out in Section~\ref{sec:intro}, energy efficiency is of great importance to smart wrist-worn devices. Existing methods fail to provide an ideal solution for common smart watches or wristbands. Furthermore, existing models for OSA detection based on RR-interval signals lack the consideration of physiological properties of human respiration, e.g., the dynamics of breathing frequencies.
In light of the feasibility of \cite{surrel2016low}, we assume respiration signals are coupled with RR-interval signals, and accordingly design a multi-frequency feature extraction model which automatically generates pertinent features and produces classification output on discontinuous input signals.
\begin{figure}[!t]
\centerline{\includegraphics[scale=0.38]{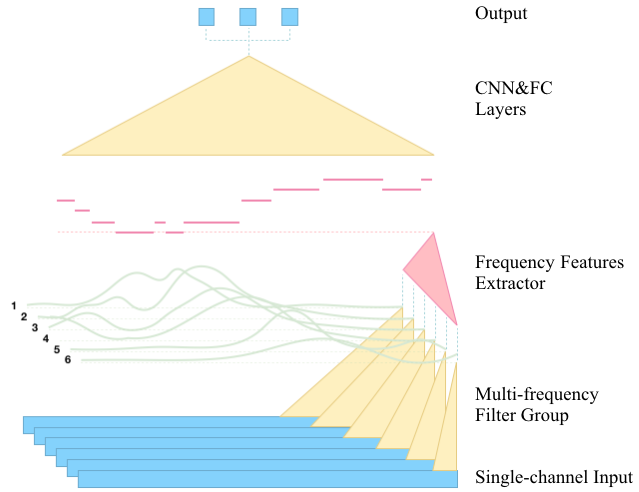}}
\vspace{-0.2cm}
\caption{The workflow of FENet.}
\vspace{-0.4cm}
\label{overview1}
\end{figure}
\section{Methodology}
In this section, we present our proposed solution to OSA detection, namely the frequency extraction network (FENet).
\subsection{Preliminaries}
To start with, we first present the definitions of RR-interval sequence and apnea label sequence used in our model.

\textbf{Definition 1: RR-interval Sequence.} 
FENet takes a sequence of RR-interval records as its input, which is collected by pulse sensors. Given a timestamp $\tau$, a single RR-interval record $I(\tau)$ is defined as follows:
\begin{equation}
I(\tau) = t(R_{i}) - t(R_{i-1}) ,\quad t(R_{i-1}) \leq \tau < t(R_{i}),
\label{eq1}
\end{equation}
where $t(\cdot)$ returns the timestamp of a pulse $R$, and $R_{i-1}$ and $R_i$ are two consecutive pulses around time $\tau$. All timestamps are in seconds. For each patient $p$, we can divide her/his full pulse sequence into $n_p$ one-minute intervals. Then, for every second in the $i$-th interval ($1\leq i \leq n_p$), we can obtain one record $I(t)$, thus obtaining a $60$-dimensional feature vector, denoted by $\textbf{x}_i = [I(60i-59), I(60i-58), ..., I(60i)]$. Ultimately, by stacking all $n_p$ feature vectors, we can represent the RR-interval sequence for patient $p$ using a feature matrix $\mathcal{X}_p^{full} = \{\textbf{x}_1,\textbf{x}_2, ...,\textbf{x}_{n_p}\}$. Note that in the rest of this paper, we refer to each one-minute interval, i.e., $\textbf{x}_i$ as an \textbf{\textit{epoch}}.

\textbf{Definition 2: Discontinuous RR-interval Sequence.} As discussed in Section \ref{sec:intro}, a major advantage of FENet is its capability of operating on discontinuous input sequence. Essentially, the discontinuous RR-interval is a downsampled version of $\mathcal{X}_p^{full}$ by skipping several epochs between two feature vectors. In our paper, the skipping step size is set to $2$ epochs, leading to only $\frac{n_p}{3}$ epochs to be collected by the pulse sensor. The resulted discontinuous RR-interval sequence is represented by $\mathcal{X}_p = \{\textbf{x}_2,\textbf{x}_5, ...,\textbf{x}_{n_p-1}\}$, where we skip the previous and subsequent epochs of each feature $\textbf{x}_i\in \mathcal{X}_p$. 

\textbf{Definition 3: Apnea Label Sequence}. A binary label $a_i \in \{ 0, 1\}$ is used to denote whether there is an apnea in the $i$-th epoch, where we set $a_i = 1$ for positive and $a_i = 0$ for negative. For patient $p$, we organize the nested apnea vector sequence $\mathcal{Y}_p$ in correspondence to the format of discontinuous $\mathcal{X}_p$, i.e., $\mathcal{Y}_p = \{\textbf{y}_2, \textbf{y}_5, ..., \textbf{y}_{n_p-1}\}$. In $\mathcal{Y}_p$, each $y_i = \{a_{i-1}, a_{i}, a_{i+1}\}$ contains three labels indicating the apnea status for the current, previous, and subsequent epochs. Then we can easily get continuous apnea label sequence $\mathcal{A}_p = \{a_1, a_2, ..., a_{n_p}\}$ by unfolding $\mathcal{Y}_p$.

Finally, we define the OSA detection problem as follows.

\textbf{Problem Definition: OSA Detection on Discontinuous RR-interval Signals.} For each user $p$, given her/his discontinuous RR-interval sequence $\mathcal{X}_p$, our goal is to learn an OSA detection model to map $\mathcal{X}_q$ to the nested apnea vector sequence $\mathcal{Y}_q$.

\begin{figure}[t]
\centering
\subfigure[]{
\includegraphics[scale=0.31]{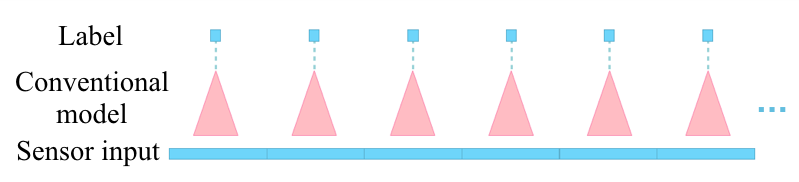} 
\label{fig:standard_state}
}
\quad
\subfigure[]{
\includegraphics[scale=0.31]{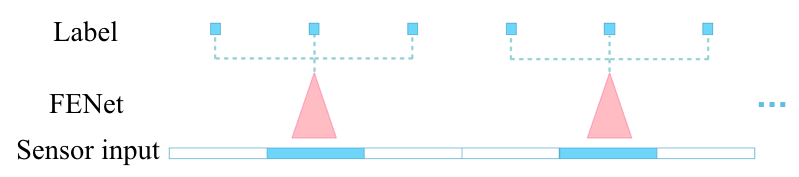}
\label{fig:energy_saving_state}
}
\quad
\vspace{-0.35cm}
\caption{Comparison between FENet and conventional models. The state of the sensor is indicated by the color in sensor input tracks (blue: ON, white: OFF). (a) shows the working process of conventional models. The sensor has to work all the time so that continuous labels can be predicted. (b) shows the working process of FENet, where one epoch of input from the sensor can generate three labels.}
\vspace{-0.65cm}
\label{fig:energy_saving_mode}
\end{figure}

\subsection{An Overview of FENet}
The workflow of Frequency Extraction Network (FENet) is demonstrated in  Fig.~\ref{overview1}. Since apnea has the symptom of a weakened or vanishing of breath, the intuition behind FENet is to fully capture the underlying respiratory processes in RR-interval signals. FENet takes one interval vector $\textbf{x}_i$ as input and then generates an implicit representation of respiration with a group of multi-frequency filters  and a frequency feature extractor. Afterwards, convolutional layers and fully connected layers are adopted to produce three independent classification results $y_i = \{a_{i-1}, a_i, a_{i+1}\}$. This design significantly reduces up to 67\% of the operation time needed from the pulse sensors on wearable devices, which is illustrated in Figure~\ref{fig:energy_saving_mode}. In what follows, we introduce the design of each key component of FENet in detail.
\begin{figure}[t]
\centering
\subfigure[]{
\includegraphics[scale=0.24]{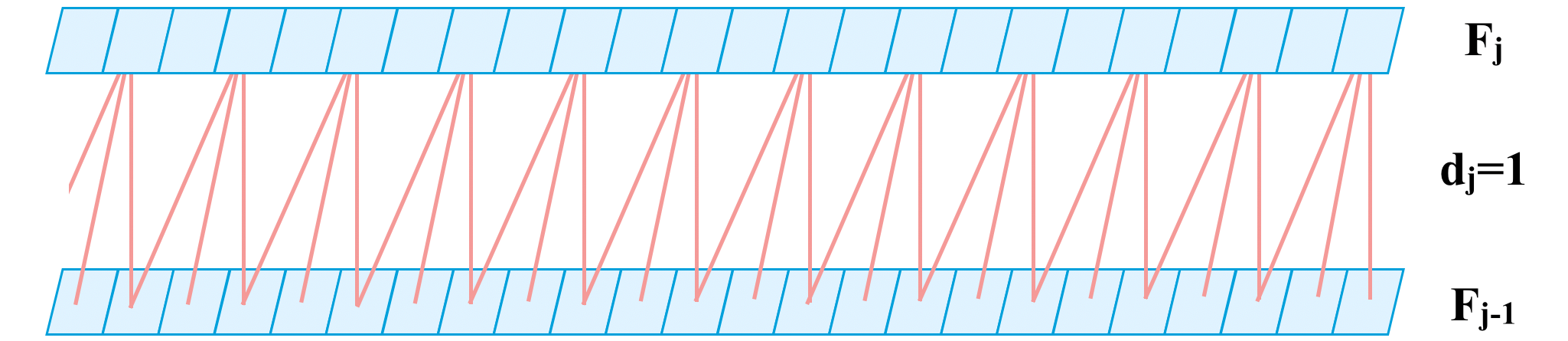}
}
\quad 
\subfigure[]{
\includegraphics[scale=0.24]{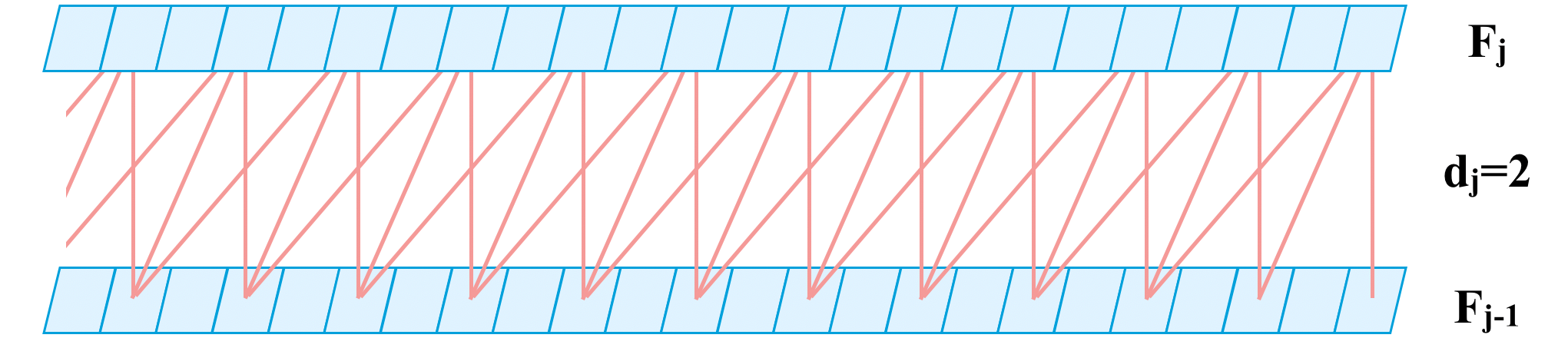}
}
\quad
\subfigure[]{
\includegraphics[scale=0.24]{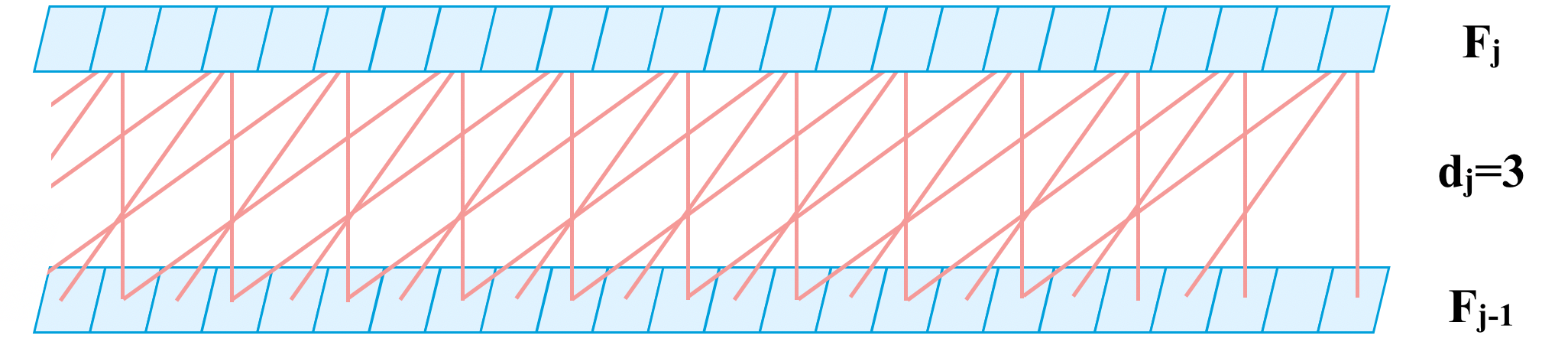}
}
\quad
\vspace{-0.5cm}
\caption{The larger the value of $d_j$, the more global context can be captured by one neuron in $F_j$, which enables FENet to extract similar events with different size (frequency).
}
\label{fig:dilated_conv}
\vspace{-0.6cm}
\end{figure}
\subsection{Multi-frequency Dilated Convolutional Neural Network}
Motivated by the proven effectiveness of CNNs in signal processing \cite{baseline_CNN},  we build the feature extraction layer of FENet upon the CNN architecture. First, let us recall the standard definition of a convolution operation:
\begin{equation}
(\textbf{F} * \textbf{k})(n) = \sum_{s+t=n} \textbf{F}(s)\textbf{k}(t),
\label{eq:convolution1}
\end{equation}
where $\textbf{F}$ denotes the raw input feature (i.e., vector $\textbf{x}_i$ in our case), $\textbf{k}$ is the filter (a.k.a. convolution kernel), and $\textbf{F}*\textbf{k}$ denotes the generated feature after performing convolution with $\textbf{k}$ on $\textbf{F}$. $n$, $s$ and $t$ are respectively the indexes of elements in feature $\textbf{F}*\textbf{k}$, input $\textbf{F}$ and kernel $\textbf{k}$. For every human, the size of each breath changes constantly, which leads to fluctuations in the frequency of human respiration. Correspondingly, we propose to comprehensively extract such multi-frequency feature with CNNs. Though this notion is previously termed as multi-resolution feature modelling in existing work \cite{b23}, the proposed multi-resolution CNN only focuses on signals that contain heterogeneous sources and carry rich semantics, e.g., a mixture of audio sources. However, when dealing with the homogeneous RR-interval signals in OSA detection, it can hardly discover sufficient frequency-based patterns that are subtle yet important. To derive a fine-grained CNN-based model that can thoroughly learn the multi-frequency features from raw RR-interval signals, we build our model upon an 1-dimensional (1-D) dilated convolutional layer \cite{dilatedconvolutions} with a group of multi-frequency filters. Denoted as $\textbf{F}*_d \textbf{k}$, the dilated convolution operation proceeds as the following:
\begin{equation}
(\textbf{F} *_d \textbf{k})(n) = \sum_{s + dt = n} \textbf{F}(s)\textbf{k}(t),
\label{eq:convolution2}
\end{equation}
where $d$ is the dilation coefficient, and $\textbf{k}$ is an $1\times 3$ convolution filter determined from $\{1\times 3, 1\times 5, 1\times 7\}$ via hyperparameter grid search. When $d=1$, Eq.(\ref{eq:convolution2}) enables the standard convolution with a $3$-dimensional receptive field for 1-D convolution, while $d>1$ will enlarge the receptive field to $(2d+1)$-dimensional. Hence, the feature in the $j$-th layer of our multi-frequency dilated CNN can be defined as:
\begin{equation}
\textbf{F}_{j} = \textbf{F}_{j-1} *_{d_j} \textbf{k}_j.
\label{filter1}
\end{equation}
\begin{figure}[]
\centering
\subfigure[]{
\begin{minipage}[t]{0.5\linewidth}
\centering
\includegraphics[scale=0.32]{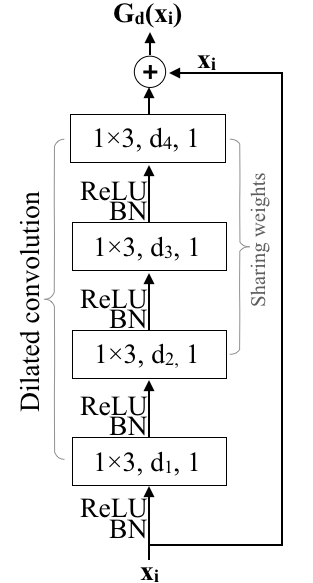}
%\caption{fig1}
\end{minipage}%
\label{fig:residual}
}%
\subfigure[]{
\begin{minipage}[t]{0.47\linewidth}
%\centering
\includegraphics[scale=0.32]{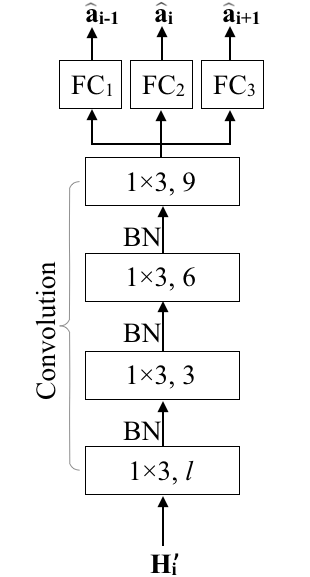}
%\caption{fig2}
\end{minipage}%
\label{fig:CNNFC}
}%
\centering
\vspace{-0.3cm}
\caption{Key components in FENet. (a) shows our proposed multi-frequency dilated convolutional network. (b) is a schematic view of the classifier utilized in FENet.}
\vspace{-0.4cm}
\label{fig:network_structure}
\end{figure}
Our multi-frequency dilated CNN works like a wavelet basis that can collect information in its corresponding frequency. Specifically, the frequency of filter $ \textbf{k}_j $ is:
\begin{equation}
freq(\textbf{k}_j) = \frac{freq_s}{d_j},
\label{freq}
\end{equation}
where $freq_s$ is the sampling frequency of input signals. In FENet, we adopt a 4-layer structure in the multi-frequency dilated CNN by default. As shown in Figure~\ref{fig:dilated_conv}, the larger the value of $d_1$, more global context can be captured from the input. Figure~\ref{fig:residual} depicts a schematic architecture of our multi-frequency dilated CNN. The filters are shared except the bottom layer, allowing FENet to automatically adapt to the amplitude of different frequencies while minimizing the computational 
complexity. Let $\textbf{F}_0=\textbf{x}_i$, then for each group of dilation coefficients $d=\{d_1, d_2, d_3, d_4\}$, we can obtain the following convolutional feature:
\begin{equation}
G_{d}(\textbf{x}_i) = \textbf{x}_i + (\textbf{x}_i *_{d_1} \textbf{k}_{1} *_{d_2} \textbf{k}_2 *_{d_3} \textbf{k}_3 *_{d_4} \textbf{k}_4),
\label{filter2}
\end{equation}
where the first term is the residual connection \cite{b24} we imposed for faster convergence, and $\textbf{k}_2 = \textbf{k}_3 = \textbf{k}_4$, as illustrated in Figure~\ref{fig:residual}. To be succinct, in Eq.(\ref{filter2}), we omit the nonlinear activation with rectified linear unit (ReLU) and batch normalization (BN) applied to each layer's input. 

\subsection{Frequency Feature Extractor}
Let $\mathcal{D}$ be the set of all coefficients $d$, then for  input $\textbf{x}_i$, we combine all $|\mathcal{D}|$ generated convolutional features to obtain a unified representation of respiratory epoch $i$:
\begin{equation}
\textbf{H}_i = \concat_{\forall d\in \mathcal{D}} G_{d}(\textbf{x}_i),
\label{eq:H}
\end{equation}
where $\concat$ denotes the iterative operator for vector concatenation that generates the feature map $\textbf{H}_i\in \mathbb{R}^{|\mathcal{D}|\times 60}$. $\textbf{H}_i$ now stacks all multi-frequency representations of epoch $i$ learned by the dilated CNN with $|\mathcal{D}|$ distinct sets of filters. To further model the intricate feature interactions between different filter groups, we additionally adopt an 1-D convolution layer with $l \leq |\mathcal{D}|$ kernels 
as the frequency feature extractor:
\begin{equation}
\textbf{H}_{i}' = CNN(\textbf{H}_i).
\label{h}
\end{equation}

As such, $\textbf{H}'_i\in \mathbb{R}^{l\times 60}$ is a compact yet expressive representation that encodes the respiratory properties within the $i$-th breathing epoch.

\subsection{OSA Detection}
With the resulted latent representation $\textbf{H}_i'$ for a single epoch, the OSA detection module in FENet is essentially a classifier for estimating three binary labels for consecutive respiratory epochs $i-1$, $i$ and $i+1$. As $\textbf{H}_i'$ is already highly representative, we adopt a simple network structure for this classifier to minimize the computational complexity. Specifically, as shown in Figure~\ref{fig:CNNFC}, a 3-layer fix-sized convolution followed by three parallel fully connected units are adopted for emitting three separate classification results: 
\begin{align}
	\widehat{\textbf{H}}_i &= CNN'(\textbf{H}_i'), \\
	\{\widehat{\textbf{a}}_{i-1},\widehat{\textbf{a}}_{i},\widehat{\textbf{a}}_{i+1}\}&= \{FC_1(\widehat{\textbf{H}}_{i-1}),FC_2(\widehat{\textbf{H}}_i),FC_3(\widehat{\textbf{H}}_{i+1})\}, \nonumber
\end{align}
where $CNN'(\cdot)$ is the convolutional part (Figure~\ref{fig:CNNFC}), $FC_1(\cdot), FC_2(\cdot), FC_3(\cdot)$ denote the fully connected units that output three 2-dimensional vectors $\widehat{a}_{i-1},\widehat{a}_{i},\widehat{a}_{i+1} \in \mathbb{R}^{2}$. We introduce batch normalization and dropout respectively in $CNN'(\cdot)$ and $FC(\cdot)$, as a form of regularization. 
Finally, we apply softmax to obtain a probability distribution over two classes (i.e., labels):
\begin{align}
	\hat{\textbf{p}}_{i'} &= [\widehat{\mathbb{P}}_{i'}(label=0),\widehat{\mathbb{P}}_{i'}(label=1)] = \textnormal{softmax}(\widehat{\textbf{a}}_{i'}),\nonumber\\
	i'&\in\{i-1, i, i+1\},
\end{align}
where the resulted $\hat{\textbf{p}}_{i-1}$,$\hat{\textbf{p}}_{i}$ and $\hat{\textbf{p}}_{i+1}$ carry the probabilities of negative (i.e., 0) and positive (i.e., 1) OSA symptoms inferred from the current epoch $\textbf{x}_i$. Correspondingly, in the inference stage, the predicted labels $\widehat{a}_{i-1}$, $\widehat{a}_{i}$ and $\widehat{a}_{i+1}$ for OSA disease is computed via:
\begin{equation}
\widehat{a}_{i'} = 
\begin{cases}
0,& \textnormal{if} \,\,\, \widehat{\mathbb{P}}_{i'}(label=0)\geq \widehat{\mathbb{P}}_{i'}(label=1)\\
1,& \textnormal{otherwise}
\end{cases},
\end{equation}
where $i'\in\{i-1, i, i+1\}$.

\subsection{Optimization Strategy}
To facilitate training for FENet, we introduce our modified cross-entropy loss to quantify the classification error on each patient $p$:
\begin{align}\label{eq:loss1}
\mathcal{L}_p = -\sum_{\forall i \in \mathcal{X}_p}&\Big{(}\lambda_1 \cdot a_{i-1}\log(\widehat{\mathbb{P}}_{i-1}(label=a_{i-1})) \nonumber\\
	 &\!\!\!\! + \lambda_2 \cdot a_{i}\log(\widehat{\mathbb{P}}_i(label=a_i)) \\ 
	 &\!\!\!\! + \lambda_3 \cdot a_{i+1}\log(\widehat{\mathbb{P}}_{i+1}(label=a_{i+1}))\Big{)}\nonumber,
\end{align}
which is them summed up for all $p\in \mathcal{P}$:
\begin{equation}
	\mathcal{L} = \sum_{\forall p\in \mathcal{P}}\mathcal{L}_p,
\end{equation}
where $\lambda_2 > 0$ is the weight assigned to the $i$-th epoch, while $\lambda_1,\lambda_3 >0$ are weights controlling the impact of the preceding and succeeding epochs, respectively. We also define $\lambda_1 + \lambda_2 + \lambda_3 = 1$. Since FENet is an end-to-end neural network, we adopt stochastic gradient descent (SGD) to optimize its parameters. To be specific, we leverage a mini-batch SGD-based optimizer, namely Adam \cite{adam} to minimize $\mathcal{L}$ until it converges.

\begin{table}[t]
\caption{Key Statistics of of PAD, UCDSAD and BestAIR.}
\vspace{-0.4cm}
\begin{center}
\begin{tabular}{c|c|c|c|c|c|c}
\hline
\multirow{2}{*}{Source} & \multirow{2}{*}{Age} &\multicolumn{3}{c}{Gender} & \multicolumn{2}{|c}{AHI}\\
\cline{3-7}
 &  & Male & Female & Unknown & $\leq$15 & $>$15 \\
\hline
PAD  & 32$\sim$56 	& 57 	& 13 & 0 &  28	&  42	\\
UCDSAD  & 28$\sim$68 	& 21 	& 4 & 0 &  12	&  13	\\
BestAIR  & 45$\sim$75 	& 53 	& 51 & 44 &  0	&  148	\\
\hline
\end{tabular}
\label{tab:DatasetInfo}
\end{center}
\vspace{-0.6cm}
\end{table}

\section{Experiment}
In this section, we conduct experiments on real-world datasets to verify the effectiveness of FENet in OSA detection.

\subsection{Datasets and Evaluation Metrics}
We gather $243$ recordings  from three data sources, namely the PhysioNet apnea-ECG dataset (PAD) \cite{b11}, University College of Dublin's sleep apnea dataset (UCDSAD) \cite{Dublin} and the Best Apnea Interventions for Research (BestAIR) \cite{BestAIR2}. PAD was released for the Cardiology Challenge 2000, and originally contained the labelled recordings from $35$ people. Recently, PAD has further released $35$ labelled recordings which were initially held out for evaluation purposes in this challenge. UCDSAD contains annotated ECG signals from $25$ adults. It includes four kinds of diagnoses: hypopnea, central sleep apnea, obstructive sleep apnea and mixed sleep apnea, where the last two types are labelled as OSA events in our experiments. BestAIR is released by the National Sleep Research Resource (NSRR) \cite{BestAIR1}, which contains 148 individuals. We list some key statistics and OSA severity information of all data sources in Table~\ref{tab:DatasetInfo}, where we use apnea-hypopnea index (AHI) to indicate the severity of OSA. An OSA case is considered normal/mild if AHI$\leq$$15$ and moderate/severe otherwise. With these data sources, we are able to obtain three following evaluation datasets:

\begin{itemize}
	\item \textbf{PAD-UCDSAD:} As both PAD and UCDSAD are relatively small and diverse in severity of OSA, we combine all $95$ recordings from two sources for performance test, termed PAD-UCDSAD.
	\item \textbf{PAD$^*$:} As all baseline methods are originally designed and optimized based on the first $35$ people's recordings in the competition, we denote the first half of PAD as PAD$^*$ and use it as an additional evaluation dataset.
	\item \textbf{BestAIR:}  In \cite{BestAIR1}, a benchmark dataset with $148$ individuals and 129,680 valid epochs was collected. This is the largest one among all three experimental datasets.
\end{itemize}

\begin{figure}[t]
\centerline{\includegraphics[scale=0.24]{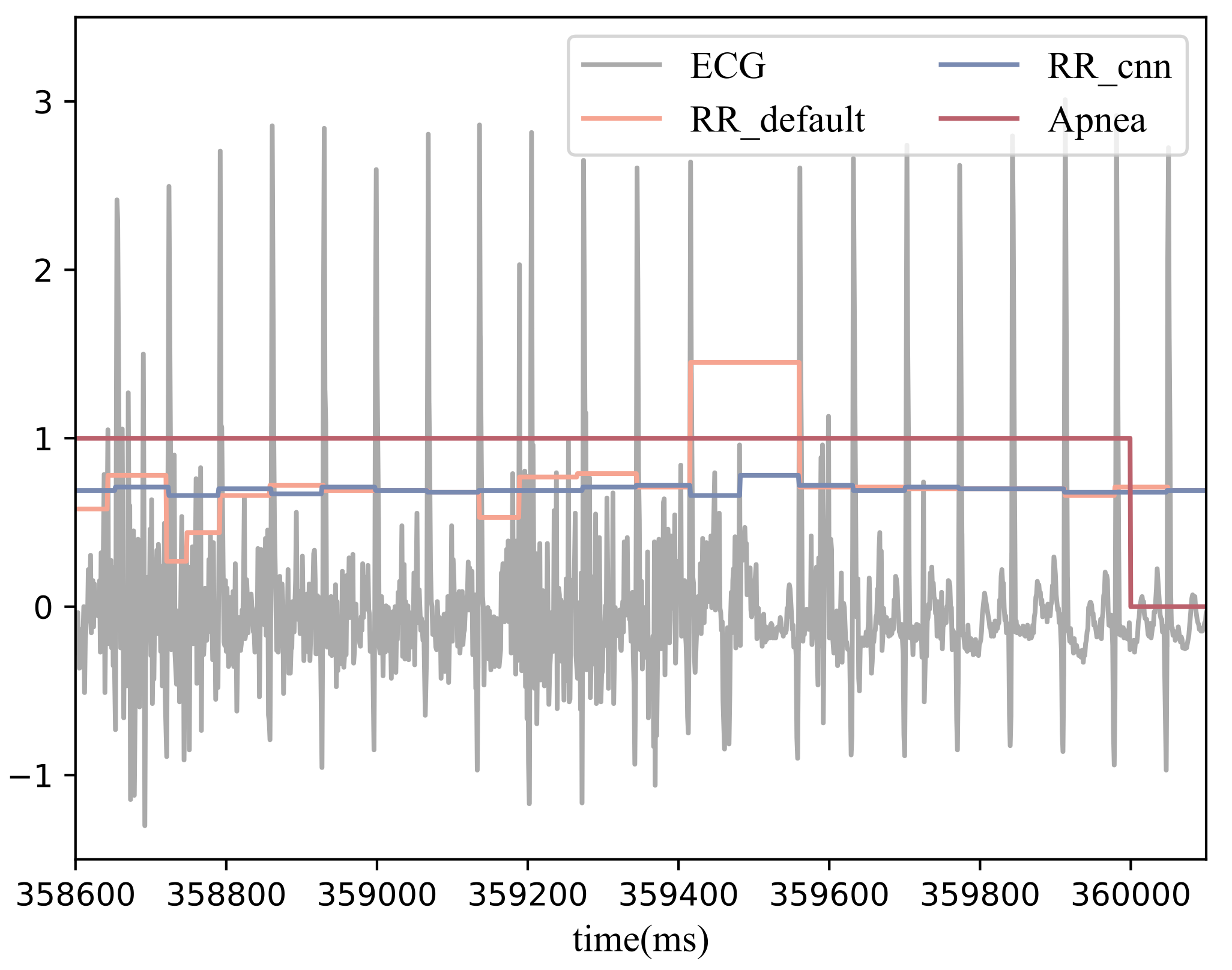}}
\vspace{-0.3cm}
\caption{A slice of one data sample from PAD$^*$. We use grey, red, orange, and blue to mark the raw ECG signals, the OSA labels, the  RR-intervals detected by the default algorithm \cite{b11}, and the RR-intervals detected by our chosen algorithm \cite{b22}.}
\label{fig:recording_slice}
\vspace{-0.5cm}
\end{figure}

For all datasets, we employ a well-established, CNN-based QRS complex detection algorithm \cite{b22} to locate every R-peak in ECG records. This offers more accurate RR-interval signals compared with the one used in \cite{b11}. The partial output of QRS detection algorithm \cite{b22} is visualized as the blue line in Figure~\ref{fig:recording_slice}, which presents better stability than the default one. We eventually get 44,279, 16,888, and 129,680 valid epochs from PAD-UCDSAD, PAD$^*$, and BestAIR, respectively.

\subsection{Baseline Methods and Hyperparameter Settings}
We compare FENet with the following RR-interval-based OSA detection models:
\begin{itemize}
\item \textbf{CNN}\cite{baseline_CNN}: It is a CNN-based model for OSA detection designed by Wang \textit{et al.}, which contains three convolutional layers, two max-pooling layers and three fully connected layers.
\item \textbf{LSTM}\cite{baseline_LSTM}: This is an end-to-end model consisting of four LSTM layers and one softmax layer. Each LSTM layer is followed by a batch normalization operation. 
\item \textbf{CRNN} \cite{b21}: This is a hybrid deep learning model that combines a 2-layer CNN with a bidirectional LSTM for OSA detection.
\end{itemize}
For FENet, apart from the convolution filter size, all optimal hyperparameters are determined via grid search. For the number of filters in the CNN-based frequency extractor, it is searched in $l\in\{1,2,3,4\}$. Also, we assume $\lambda_1$ and $\lambda_3$ in Eq.(\ref{eq:loss1}) are equally important, hence $\lambda_1 = \lambda_3 = 0.5(1-\lambda_2)$ and $\lambda_2$ is tuned in $\{0.9,0.7,0.5,0.3\}$. To simplify the search space for $|\mathcal{D}|$ sets of dilation coefficients $\{d_1, d_2, d_3, d_4\}$, we fix the dilation rate for upper layers with $\{d_2=2, d_3=4, d_4=8\}$ and vary $d_1$. As the regular respiration rate for sleeping adults is 12 to 20 breaths per minute, the lower bound of $d_1$ is 3 according to Eq.(\ref{freq}). In our experiments, we adopt four values $\{3,4,5,6\}$ for  $d_1$, hence $\mathcal{D}=\{\{d_1, d_2, d_3, d_4\}|d_1\in\{3,4,5,6\}\}$. For all our baseline methods, we directly adopt their reported optimal hyperparameter settings.

\subsection{Evaluation Protocol}
\textbf{Evaluation Metrics.} We use three popular metrics for classification tasks, namely Accuracy (Acc), Recall (Rec), Precision (Pre) and Specificity (Spe). They are defined as follows:
\begin{equation}
\begin{split}
Acc &= \frac{TP + TN}{TP + TN + FP + FN}, \,\,\, Rec = \frac{TP}{TP + FN},\\
Pre &= \frac{TP}{TP + FP}, \hspace{2.2cm}Spe = \frac{TN}{TN + FP},	
\end{split}
\end{equation}
where $TP$ is the number of correctly detected abnormal epochs (i.e., true positive cases), and $TN$ is the number of correctly detected normal epochs (i.e., true negative cases). $FP$ is the number of incorrectly detected abnormal epochs (i.e., false negative cases) and $FN$ is the number of incorrectly detected normal epochs (i.e., false negative cases). From the perspective of OSA detection, false negative cases is much more severe compared with false positive cases as higher $FN$ means more abnormal epochs are missed. Hence, our foremost aim is to reduce $FN$, and we prioritize the importance of Recall above Accuracy, Precision and Specificity. 

\subsection{Effectiveness on Continuous OSA Detection}
Because the baseline methods are all based on consecutive RR-interval signals, we firstly verify all methods' effectiveness by performing OSA detection in a continuous manner. In other words, we assume the availability of $\mathcal{X}^{full}_p$, and FENet only needs to output one label $\widehat{a}_i$ with each input $\textbf{x}_i$. Considering all baselines are originally optimized on PAD$^{*}$, we use PAD$^{*}$ as an additional evaluation dataset on top of the full dataset PAD-USDSAD. The datasets are randomly divided for training, validation, and test with the ratio of 3:1:1.

We report the performance on continuous OSA detection of all tested models on PAD$^{*}$ and PAD-UCDSAD in Table~\ref{tab:continuous}. It is clear that our FENet consistently and significantly outperforms other baseline models on four evaluation metrics. Apart from the superiority on the commonly used PAD$^*$ dataset, FENet also maintains its high detection accuracy despite the dramatic declines from all baselines, showcasing its stability and generalizability when dealing with large-scale datasets. Compared with three baselines on Recall, FENet achieves 1.08\%, 6.47\% and 1.98\% average improvement on PAD$^*$, PAD-UCDSAD and BestAIR, respectively. 

\begin{table}[t]
\caption{Continuous OSA detection results on PAD$^*$, PAD-UCDSAD and BestAIR.}
\vspace{-0.5cm}
\begin{center}
\begin{tabular}{c|c|c|c|c|c}
\hline
Dataset & Metric & CNN & LSTM & CRNN & \textbf{FENet}\\
\hline
\multirow{4}{*}{PAD$^*$} 
& Acc & 0.9710 	& 0.9665 & 0.9754 & \textbf{0.9922}  \\
& Rec & 0.9844  & 0.9791 & 0.9815 & \textbf{0.9925} \\
& Pre & 0.9823  & 0.9823 & 0.9357 & \textbf{0.9985} \\
& Spe & 0.8838  & 0.8845 & 0.8919 & \textbf{0.9902} \\
\hline
\multirow{4}{*}{PAD-UCDSAD} 
& Acc & 0.8916 	& 0.8820 & 0.8919 & \textbf{0.9663}  \\  
& Rec & 0.9561  & 0.8653 & 0.8905 & \textbf{0.9687} \\
& Pre & 0.9118  & 0.9763 & 0.9641 & \textbf{0.9866} \\
& Spe & 0.7103  & 0.9342 & 0.8961 & \textbf{0.9588} \\
\hline
\multirow{4}{*}{BestAIR}
& Acc & 0.9696 & 0.9384 & 0.9663 & \textbf{0.9846}\\
& Rec & 0.9819 & 0.9458 & 0.9741 & \textbf{0.9871} \\
& Pre & 0.9862  & 0.9891 & 0.9904 & \textbf{0.9968} \\
& Spe & 0.7244  & 0.7909 & 0.8106 & \textbf{0.9364} \\
\hline
\end{tabular}
\label{tab:continuous}
\end{center}
\vspace{-0.6cm}
\end{table}

\begin{table}[t]
\caption{Discontinuous OSA detection results on PAD-UCDSAD.}
\vspace{-0.2cm}
\begin{center}
\begin{tabular}{c|c|c|c|c|c}
\hline
Dataset & Metric & CNN & LSTM & CRNN & \textbf{FENet}\\
\hline
\multirow{4}{*}{Normal} 
& Acc & 0.8715	& 0.8515 & 0.8041 & \textbf{0.9567}  \\
& Rec & 0.9077  & 0.8501 & 0.8212 & \textbf{0.9608} \\
& Pre & 0.9216 & 0.9487 & 0.9117 & \textbf{0.9793} \\
& Spe & 0.7581  & 0.8560 & 0.7508 & \textbf{0.9457} \\
\hline
\multirow{4}{*}{Cold-start} 
& Acc & 0.7196 & 0.6915 & 0.6919 & \textbf{0.7825}  \\  
& Rec & 0.8623  & 0.7838 & 0.7824 & \textbf{0.9064} \\
& Pre & 0.7769  & 0.7905 & 0.7918 & \textbf{0.8154} \\
& Spe & 0.3385  & 0.4451 & 0.4504 & \textbf{0.4518} \\
\hline
\end{tabular}
\label{tab:discontinuous}
\end{center}
\vspace{-0.cm}
\end{table}

\subsection{Effectiveness on Discontinuous OSA Detection}
The advantage of FENet is the ability to perform prediction for three consecutive epochs using data from a single epoch. Hence, we evaluate all OSA detection methods in the discontinuous setting. The experiments are conducted on PAD-UCDSAD dataset due to its diverse OSA severity of patients (reflected by AHI). To make the baselines compatible to this setting, for each baseline, we train three independent models (without weight sharing) to emit $a_{i-1}$, $a_{i}$ and $a_{i+1}$ given input $\textbf{x}_{i}$. All hyperparameters are inherited from continuous detection experiment so there is no validation set. 

\textbf{Discussion on Overall Performance.} In this test, we use 80\% of epochs from PAD-UCDSAD to train baseline models and FENet while the rest of them are treated as test set. The results of the discontinuous detection experiments are illustrated in the first part performance under the continuous setting, they all suffer from a significant performance drop when trying to predict the labels for the missing adjacent epochs. In contrast, FENet achieves the best performance, which suggests that the dilated convolution layers are highly capable of extracting the subtle patterns from different frequency-based granularities, thus precisely inferring the apnea status of adjacent epochs. 

\textbf{Evaluating Discontinuous OSA Detection on Cold-Start Patients.}
In a real-world application, on top of the ability to cope with discontinuous RR-intervals, OSA detection systems need to serve new patients (e.g., new smart watch users), which means all epochs in the test set are from patients that have not appeared in the training set. Hence, we simulate a cold-start scenario by randomly selecting 20 people's entire records in PAD-UCDSAD, and use all their epochs for testing. The remaining records from 75 people are used for training. Since each patient (rather than each epoch) is treated as a sample here and the total number of samples turns to 95, we apply 5-fold cross validation to avoid high variances in model performance resulted from one randomly sampled test set. We list the performance of all models in Table~\ref{tab:discontinuous}. Apparently, a performance decrease can be observed from all models in the cold-start scenario due to lack of prior knowledge about the patient. However, FENet shows the smallest impact from cold-start prediction, advancing the performance of the best baseline (i.e., CNN) by 4.41\% on Recall.

\begin{figure}[t]
\centerline{\includegraphics[scale=0.18]{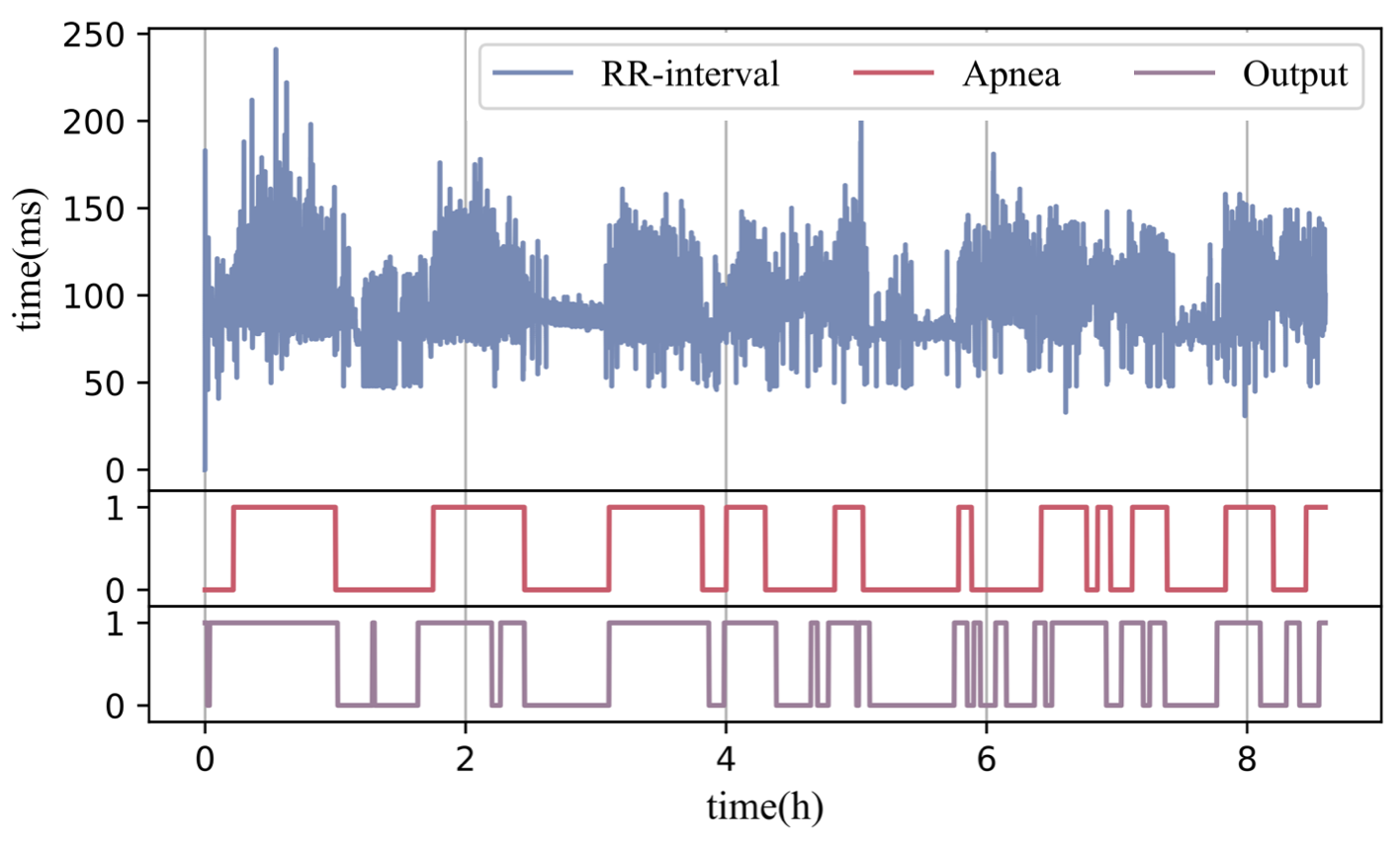}}
\vspace{-0.3cm}
\caption{A visualization of the discontinuous OSA detection result in cold-start scenario. The record is from a patient indexed A03 in the dataset. The red and purple lines are respectively the ground truth and the output of FENet.}
\label{fig:casestudy}
\vspace{-0.4cm}
\end{figure}

\textbf{Case Study.} To further demonstrate FENet's capability of accurately detecting OSA from discontinuous RR-interval signals, we visualize the detection result on a patient's full record in Figure~\ref{fig:casestudy}. It can be observed that FENet almost finds out every apnea event, while some normal epochs are labelled incorrectly. Since the difference between hypopnea and apnea is subtle, these hypopnea epochs are likely to lead to false positive predictions. 

\subsection{Impact of Hyperparameters}
\begin{figure}[b]
\centering
\subfigure[]{
\begin{minipage}[t]{0.5\linewidth}
%\centering
\includegraphics[scale=0.17]{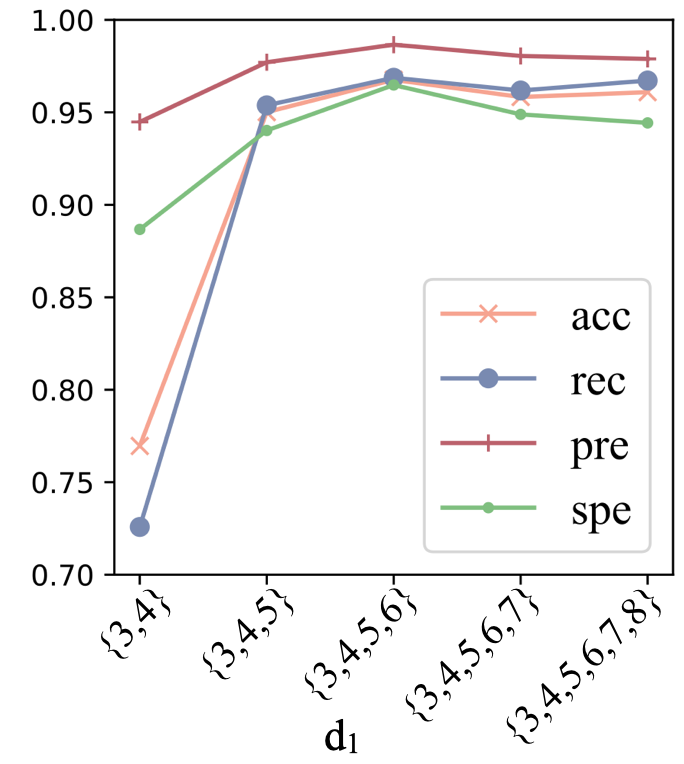}
%\caption{fig1}
\end{minipage}%
\label{fig:hyper1}
}%
\subfigure[]{
\begin{minipage}[t]{0.5\linewidth}
%\centering
\includegraphics[scale=0.17]{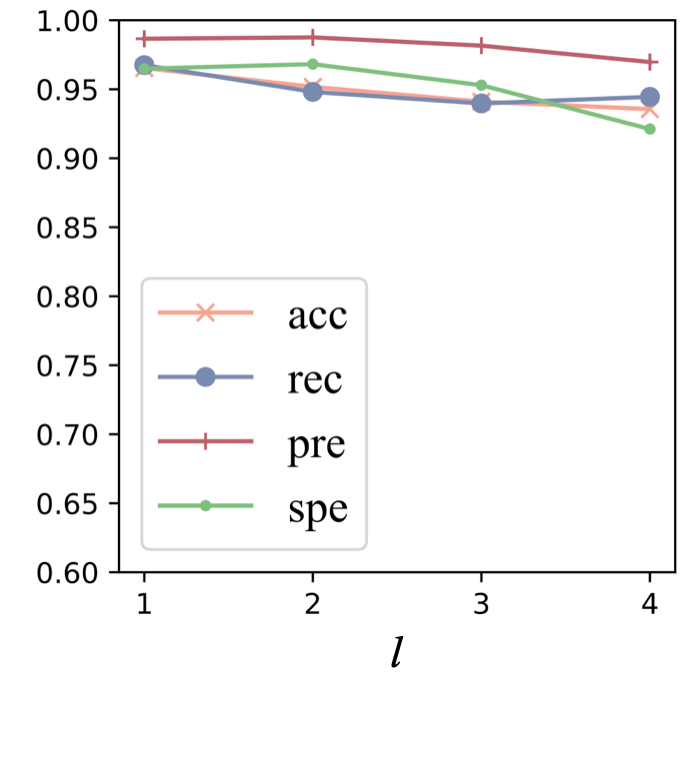}
%\caption{fig2}
\end{minipage}%
\label{fig:hyper2}
}%
\centering
\vspace{-0.3cm}
\caption{The impact of hyperparameters in FENet. (a) and (b) reflect the fluctuation of performance while varying $|\mathcal{D}|$ and $l$, respectively.}
\label{fig:hyper}
\vspace{0cm}
\end{figure}

In this study, we investigate the effect of the key hyperparameters, namely $d_1$ and $l$, on the performance of FENet. The results are depicted in Figure~\ref{fig:hyper}.
For $d_1$, we fix $l = 1$ and gradually enlarge the set of $d_1$ from $\{3,4\}$ to $\{3,4,5,6,7,8\}$. For all three metrics, the best performance can be obtained when $d_1$ has four values $\{3,4,5,6\}$, while using only two dilated filters $d_1\in\{3,4\}$ leads to an obvious performance drop, especially for Recall and Accuracy. 
Intuitively, people breath slower during asleep, hence filter in higher frequencies (lower $d_1$) may miss some important respiratory patterns. With the inclusion of $d_1=5$, FENet extracts information with around 0.2 Hz frequency from the input, which is proven beneficial for the detection performance. 

For $l$, we use four dilated filters $d_1\in\{3,4,5,6\}$, and vary it from $1$ to $4$. Generally, FENet benefits from a smaller $l$, as too many deep layers in the frequency feature extractor will bring excessive parameters, thus making the network harder to train and more vulnerable to overfitting issues. As the impact of $l$ is relatively small, FENet is deemed insensitive to $l$. 

\begin{figure}[t]
\vspace{-0.5cm}
\centering
\subfigure[]{
\begin{minipage}[t]{0.5\linewidth}
%\centering
\includegraphics[scale=0.17]{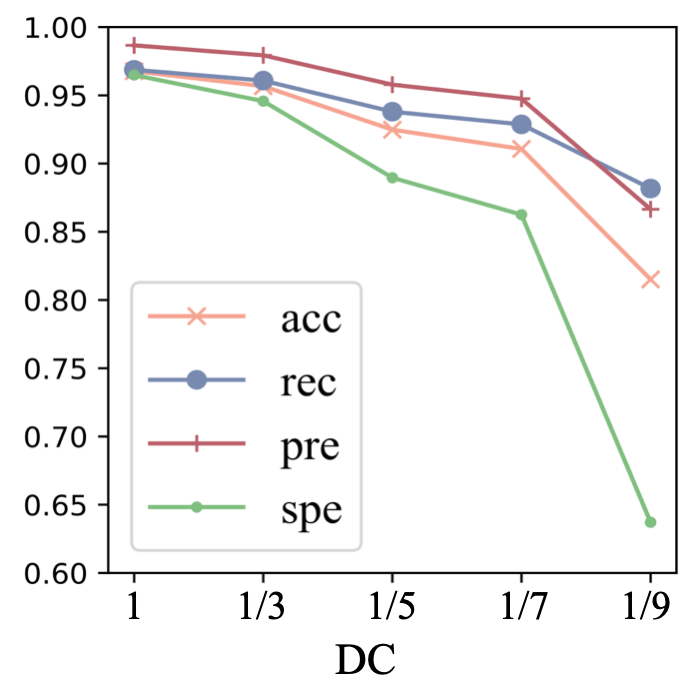}
%\caption{fig1}
\end{minipage}%
\label{fig:dutycycle}
}%
\subfigure[]{
\begin{minipage}[t]{0.5\linewidth}
%\centering
\includegraphics[scale=0.17]{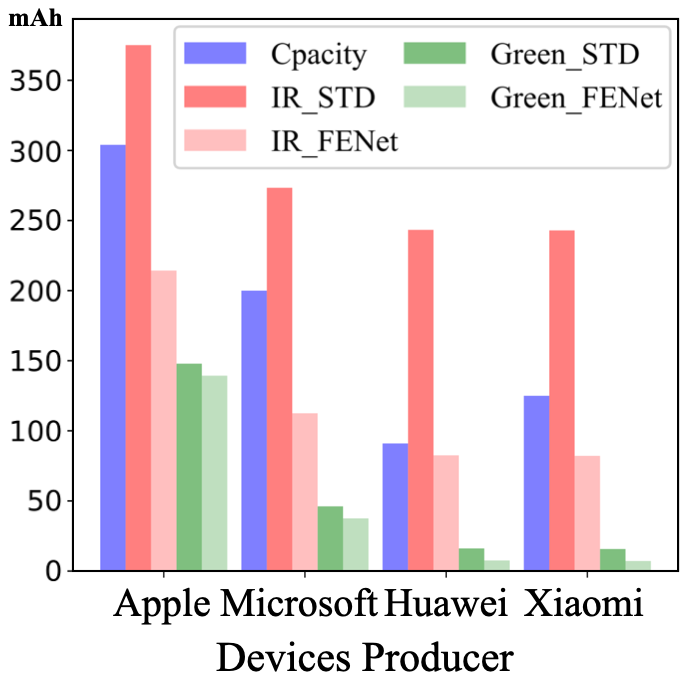}
\end{minipage}
\vspace{-1cm}
\label{fig:capacity}
}%
\centering
\vspace{-0.3cm}
\caption{(a) The performance of FENet when duty cycle decreases from 1 to $\frac{1}{9}$. (b) Energy consumption of four devices for an 8-hour OSA monitoring in different conditions. Green and IR denote two sensor configurations, while STD and FENet respectively denote the standard duty cycle adopted by baseline methods and the lowered duty cycle in FENet.}
\label{fig:energy}
\vspace{-0.55cm}
\end{figure}
\subsection{Analysis on Energy Efficiency \label{sec:energy}}
\textbf{The Impact of Duty Cycle.} Additionally, we investigate how FENet behaves if we would like to engage the PPG sensor for a shorter time for RR-interval collection. This is achieved by reducing the duty cycle (DC) of the sensor, e.g., a duty cycle of $\frac{1}{9}$ means that FENet needs to predict the labels for $9$ time steps $\{\widehat{a}_{i-4},\widehat{a}_{i-3},...,\widehat{a}_{i+3},\widehat{a}_{i+4}\}$ given each $\textbf{x}_i$, and $\frac{1}{3}$ corresponds to the default duty cycle of FENet. We report the experimental results with DC$\in\{1, \frac{1}{3}, \frac{1}{5}, \frac{1}{7}, \frac{1}{9}\}$ in Figure~\ref{fig:dutycycle}. Note that the number of FC blocks for emitting the output labels also changes accordingly. Though the sensor can theoretically save more energy with lower DC, unsatisfying performance is observed when DC is below $\frac{1}{5}$.

\textbf{Energy Consumption Simulation.} FENet reduces the energy cost of sensors, which is very useful for devices whose data collectors (i.e., PPG sensors in our case) are the primary energy consumer. In this section, we investigate the energy consumption by simulating the working condition of smart wrist-worn devices. For FENet, with the default PPG sensor duty cycle of $\frac{1}{3}$, $\frac{2}{3}$ energy can be conserved in ideal conditions if no energy consumption is incurred by other components. Given more complex real-life scenarios where the background energy consumption is greater than $0$, we investigate the energy consumption of some common smart wrist-worn devices. The devices analyzed are Apple Watch Series 6, Microsoft Band 2, Huawei Band 4 and Mi Smart Band 5. Since the specification of the PPG sensor equipped on each device is commonly unavailable from manufacturers, we select two PPG sensor modules with publicly available data sheets that are designed for wearable devices to simulate the working condition of sensors in target devices.

As shown in Table \ref{tab:Current}, we can easily obtain official information about each device's battery capacity and standard standby time, which can be used to calculate the background current $I_{BG}$ by dividing the battery power by the standby time. We then select two representative PPG sensors. The first one is BH1790GLC\footnote{https://fscdn.rohm.com/en/products/databook/datasheet/ic/sensor/\\pulse\_wave/bh1790glc-e.pdf}, which has two green LEDs and is characterized by its low power consumption. The other is OB1203\footnote{https://www.idt.com/us/en/document/dst/ob1203-datasheet}, a cutting-edge sensor module that can detect pulse waves via infra-red (IR) light. We use Green and IR to denote two sensors, respectively. According to the specifications, the working current $I_{PPG}$ of Green and IR sensors are respectively $1.6$ mA and $30$ mA. Then, assuming all devices are powered by the same voltage, we can calculate the electricity consumption (measured by mAh) for both standard and FENet's conditions in $T$ hours, i.e., $C_{STD}= (I_{PPG}+I_{BG})T$ and $C_{FENet}= (\frac{1}{3}I_{PPG}+I_{BG})T$. We visualize the electricity consumptions of an 8-hour overnight monitoring of four devices using Green/IR sensor in Figure~\ref{fig:capacity}. The battery capacity is also reported, showing that existing methods' energy consumption exceeds the battery capacity when IR sensor is used in its full duty cycle. The results suggest that, on devices with different battery and PPG sensor configurations, FENet significantly cuts down the energy consumption by allowing the PPG sensor to operate with $\frac{1}{3}$ duty cycle. Also, it enables all devices to support overnight monitoring. More importantly, as shown in Table \ref{tab:discontinuous}, the design of FENet makes it consistently outperform existing full duty cycle methods that are adapted to discontinuous OSA detection.
\begin{table}[t]
\caption{Background Current of Four Wearable Devices.}
\vspace{-0.4cm}
\begin{center}
\begin{tabular}{c|c|c|c}
\hline
 Model
 & Capacity
 & Run Time
 & $I_{BG}$\\
\hline
Apple Watch Series 6\tablefootnote{https://www.apple.com/watch/battery/}  & 304 mAh & 18 h	 & 16.89 mA \\
Microsoft Band 2\tablefootnote{https://en.wikipedia.org/wiki/Microsoft\_Band\_2} & 200 mAh& 48 h & 4.17 mA  	\\
Huawei Band 4\tablefootnote{https://consumer.huawei.com/au/wearables/band4/specs/}& 91 mAh & 216 h & 0.42 mA	\\
Mi Smart Band 5\tablefootnote{https://www.mi.com/global/mi-smart-band-5/}& 125 mAh & 336 h & 0.37 mA	\\
\hline
\end{tabular}
\label{tab:Current}
\end{center}
\vspace{-0.6cm}
\end{table}
\section{Conclusion}
In this paper, we propose FENet, a novel model that extracts features from different frequencies of the input RR-interval signals to perform OSA detection in an energy-efficient manner. We constructed a dilated convolutional neural network with a set of filters for different frequency bands. As such, FENet adapts to frequencies close to human breath and accordingly generates expressive representation of a respiration epoch. The representation is subsequently processed by one convolutional layer block and three fully connected layer blocks to produce final detection results for both the current epoch and the two adjacent epochs that are missing. In our future work, we would like to incorporate human-in-the-loop scheme into FENet to make it able to generate accurate predictions in a timely manner.

\balance
\section*{Acknowledgment}
This work was supported by ARC Discovery Project (GrantNo.DP190101985).

\ifCLASSOPTIONcaptionsoff
	\newpage
\fi

%\bibliographystyle{IEEEtran} 
%\bibliography{references} 

\begin{thebibliography}{10}
\providecommand{\url}[1]{#1}
\csname url@samestyle\endcsname
\providecommand{\newblock}{\relax}
\providecommand{\bibinfo}[2]{#2}
\providecommand{\BIBentrySTDinterwordspacing}{\spaceskip=0pt\relax}
\providecommand{\BIBentryALTinterwordstretchfactor}{4}
\providecommand{\BIBentryALTinterwordspacing}{\spaceskip=\fontdimen2\font plus
\BIBentryALTinterwordstretchfactor\fontdimen3\font minus
  \fontdimen4\font\relax}
\providecommand{\BIBforeignlanguage}[2]{{%
\expandafter\ifx\csname l@#1\endcsname\relax
\typeout{** WARNING: IEEEtran.bst: No hyphenation pattern has been}%
\typeout{** loaded for the language `#1'. Using the pattern for}%
\typeout{** the default language instead.}%
\else
\language=\csname l@#1\endcsname
\fi
#2}}
\providecommand{\BIBdecl}{\relax}
\BIBdecl

\bibitem{b26}
M.~J. Sateia, ``International classification of sleep disorders,''
  \emph{Chest}, vol. 146, no.~5, pp. 1387--1394, 2014.

\bibitem{b4}
R.~B. Berry, R.~Budhiraja, D.~J. Gottlieb, D.~Gozal, C.~Iber, V.~K. Kapur,
  C.~L. Marcus, R.~Mehra, S.~Parthasarathy, S.~F. Quan \emph{et~al.}, ``Rules
  for scoring respiratory events in sleep: update of the 2007 aasm manual for
  the scoring of sleep and associated events: deliberations of the sleep apnea
  definitions task force of the american academy of sleep medicine,''
  \emph{Journal of clinical sleep medicine}, vol.~8, no.~5, pp. 597--619, 2012.

\bibitem{b12}
F.~Mendonca, S.~S. Mostafa, A.~G. Ravelo-Garc{\'\i}a, F.~Morgado-Dias, and
  T.~Penzel, ``A review of obstructive sleep apnea detection approaches,''
  \emph{IEEE journal of biomedical and health informatics}, vol.~23, no.~2, pp.
  825--837, 2018.

\bibitem{b1}
J.~Espiritu, ``Health consequences of obstructive sleep apnea,'' \emph{J Sleep
  Disord Ther}, vol.~8, no.~5, pp. 1--17, 2019.

\bibitem{b2}
A.~Shojaee and V.~Mohsenin, \emph{Obstructive Sleep Apnea Increases the Risk of
  Pulmonary Hypertension Independent of Relevant Risk Factors}.\hskip 1em plus
  0.5em minus 0.4em\relax American Thoracic Society, 2020.

\bibitem{b3}
R.~Schulz, F.~Bischof, W.~Galetke, H.~Gall, J.~Heitmann, A.~Hetzenecker,
  M.~Laudenburg, T.~J. Magnus, G.~Nilius, C.~Priegnitz \emph{et~al.}, ``Cpap
  therapy improves erectile function in patients with severe obstructive sleep
  apnea,'' \emph{Sleep medicine}, vol.~53, pp. 189--194, 2019.

\bibitem{b11}
Y.~Ichimaru and G.~Moody, ``Development of the polysomnographic database on
  cd-rom,'' \emph{Psychiatry and clinical neurosciences}, vol.~53, no.~2, pp.
  175--177, 1999.

\bibitem{b5}
A.~Sabil, M.~Glos, A.~G{\"u}nther, C.~Sch{\"o}bel, C.~Veauthier, I.~Fietze, and
  T.~Penzel, ``Comparison of apnea detection using oronasal thermal airflow
  sensor, nasal pressure transducer, respiratory inductance plethysmography and
  tracheal sound sensor,'' \emph{Journal of Clinical Sleep Medicine}, vol.~15,
  no.~2, pp. 285--292, 2019.

\bibitem{b6}
J.~L. Goodwin, P.~L. Enright, K.~L. Kaemingk, G.~M. Rosen, W.~J. Morgan, R.~F.
  Fregosi, and S.~F. Quan, ``Feasibility of using unattended polysomnography in
  children for research‚ A report of the tucson children's assessment of sleep
  apnea study (tucasa),'' \emph{Sleep}, 2001.

\bibitem{b7}
M.~H. Sanders, B.~K. Lind, S.~F. Quan, D.~J. Iber Conrad~Gottlieb, W.~H.
  Bonekat, D.~M. Rapoport, P.~L. Smith, and J.~P. Kiley, ``Methods for
  obtaining and analyzing unattended polysomnography data for a multicenter
  study,'' \emph{Sleep}, vol.~21, no.~7, pp. 759--767, 1998.

\bibitem{b10}
X.~Liang, X.~Qiao, and Y.~Li, ``Obstructive sleep apnea detection using
  combination of cnn and lstm techniques,'' in \emph{ITAIC}.\hskip 1em plus
  0.5em minus 0.4em\relax IEEE, 2019, pp. 1733--1736.

\bibitem{b13}
R.~Lin, R.-G. Lee, C.-L. Tseng, H.-K. Zhou, C.-F. Chao, and J.-A. Jiang, ``A
  new approach for identifying sleep apnea syndrome using wavelet transform and
  neural networks,'' \emph{Biomedical Engineering: Applications, Basis and
  Communications}, 2006.

\bibitem{b14}
A.~H. Khandoker, J.~Gubbi, and M.~Palaniswami, ``Automated scoring of
  obstructive sleep apnea and hypopnea events using short-term
  electrocardiogram recordings,'' \emph{IEEE Transactions on Information
  Technology in Biomedicine}, vol.~13, no.~6, pp. 1057--1067, 2009.

\bibitem{b15}
V.~P. Rachim, G.~Li, and W.-Y. Chung, ``Sleep apnea classification using
  ecg-signal wavelet-pca features,'' \emph{Bio-medical materials and
  engineering}, vol.~24, no.~6, pp. 2875--2882, 2014.

\bibitem{b16}
A.~Smruthy and M.~Suchetha, ``Real-time classification of healthy and apnea
  subjects using ecg signals with variational mode decomposition,'' \emph{IEEE
  sensors journal}, vol.~17, no.~10, pp. 3092--3099, 2017.

\bibitem{b17}
A.~Quiceno-Manrique, J.~Alonso-Hernandez, C.~Travieso-Gonzalez,
  M.~Ferrer-Ballester, and G.~Castellanos-Dominguez, ``Detection of obstructive
  sleep apnea in ecg recordings using time-frequency distributions and dynamic
  features,'' in \emph{EMBC}.\hskip 1em plus 0.5em minus 0.4em\relax IEEE,
  2009, pp. 5559--5562.

\bibitem{b18}
C.~Zywietz, V.~Von~Einem, B.~Widiger, and G.~Joseph, ``Ecg analysis for sleep
  apnea detection,'' \emph{Methods of information in medicine}, vol.~43,
  no.~01, pp. 56--59, 2004.

\bibitem{b19}
C.~Travieso, J.~Alonso, M.~del P.~Ba{\~n}os, J.~T.~Rivas, and K.~L.~Ipi{\~n}a,
  ``Automatic apnea identification by transformation of the cepstral domain,''
  \emph{Cognitive Computation}, vol.~5, no.~4, pp. 558--565, 2013.

\bibitem{deeplearning}
Y.~LeCun, Y.~Bengio, and G.~Hinton, ``Deep learning,'' \emph{nature}, vol. 521,
  no. 7553, pp. 436--444, 2015.

\bibitem{b20}
M.~Cheng, W.~J. Sori, F.~Jiang, A.~Khan, and S.~Liu, ``Recurrent neural network
  based classification of ecg signal features for obstruction of sleep apnea
  detection,'' in \emph{2017 IEEE International Conference on Computational
  Science and Engineering and International Conference on Embedded and
  Ubiquitous Computing}, vol.~2, 2017, pp. 199--202.

\bibitem{b21}
X.~Liang, X.~Qiao, and Y.~Li, ``Obstructive sleep apnea detection using
  combination of cnn and lstm techniques,'' in \emph{ITAIC}.\hskip 1em plus
  0.5em minus 0.4em\relax IEEE, 2019, pp. 1733--1736.

\bibitem{surrel2016low}
G.~Surrel, F.~Rinc{\'o}n, S.~Murali, and D.~Atienza, ``Low-power wearable
  system for real-time screening of obstructive sleep apnea,'' in \emph{2016
  IEEE Computer Society Annual Symposium on VLSI}, 2016, pp. 230--235.

\bibitem{surrel2018online}
G.~Surrel, A.~Aminifar, F.~Rinc{\'o}n, S.~Murali, and D.~Atienza, ``Online
  obstructive sleep apnea detection on medical wearable sensors,'' \emph{IEEE
  transactions on biomedical circuits and systems}, vol.~12, no.~4, pp.
  762--773, 2018.

\bibitem{baseline_CNN}
X.~Wang, M.~Cheng, Y.~Wang, S.~Liu, Z.~Tian, F.~Jiang, and H.~Zhang,
  ``Obstructive sleep apnea detection using ecg-sensor with convolutional
  neural networks,'' \emph{Multimedia Tools and Applications}, vol.~79, no.~23,
  pp. 15\,813--15\,827, 2020.

\bibitem{chen2020sequence}
T.~Chen, H.~Yin, Q.~V.~H. Nguyen, W.-C. Peng, X.~Li, and X.~Zhou,
  ``Sequence-aware factorization machines for temporal predictive analytics,''
  in \emph{ICDE}, 2020, pp. 1405--1416.

\bibitem{chen2018tada}
T.~Chen, H.~Yin, H.~Chen, L.~Wu, H.~Wang, X.~Zhou, and X.~Li, ``Tada: trend
  alignment with dual-attention multi-task recurrent neural networks for sales
  prediction,'' in \emph{ICDM}, 2018, pp. 49--58.

\bibitem{KDD19Streaming}
\BIBentryALTinterwordspacing
L.~Guo, H.~Yin, Q.~Wang, T.~Chen, A.~Zhou, and N.~Quoc Viet~Hung, ``Streaming
  session-based recommendation,'' in \emph{SIGKDD}, ser. KDD '19.\hskip 1em
  plus 0.5em minus 0.4em\relax New York, NY, USA: Association for Computing
  Machinery, 2019, p. 1569–1577. [Online]. Available:
  \url{https://doi.org/10.1145/3292500.3330839}
\BIBentrySTDinterwordspacing

\bibitem{WWW20Next}
\BIBentryALTinterwordspacing
Q.~Wang, H.~Yin, T.~Chen, Z.~Huang, H.~Wang, Y.~Zhao, and N.~Q. Viet~Hung,
  ``Next point-of-interest recommendation on resource-constrained mobile
  devices,'' in \emph{Proceedings of The Web Conference 2020}, ser. WWW
  '20.\hskip 1em plus 0.5em minus 0.4em\relax New York, NY, USA: Association
  for Computing Machinery, 2020, p. 906–916. [Online]. Available:
  \url{https://doi.org/10.1145/3366423.3380170}
\BIBentrySTDinterwordspacing

\bibitem{baseline_LSTM}
M.~Cheng, W.~J. Sori, F.~Jiang, A.~Khan, and S.~Liu, ``Recurrent neural network
  based classification of ecg signal features for obstruction of sleep apnea
  detection,'' in \emph{2017 IEEE International Conference on Computational
  Science and Engineering (CSE) and IEEE International Conference on Embedded
  and Ubiquitous Computing (EUC)}, vol.~2.\hskip 1em plus 0.5em minus
  0.4em\relax IEEE, 2017, pp. 199--202.

\bibitem{b23}
E.~M. Grais, H.~Wierstorf, D.~Ward, and M.~D. Plumbley, ``Multi-resolution
  fully convolutional neural networks for monaural audio source separation,''
  in \emph{International Conference on Latent Variable Analysis and Signal
  Separation}.\hskip 1em plus 0.5em minus 0.4em\relax Springer, 2018, pp.
  340--350.

\bibitem{dilatedconvolutions}
F.~Yu and V.~Koltun, ``Multi-scale context aggregation by dilated
  convolutions,'' \emph{arXiv preprint arXiv:1511.07122}, 2015.

\bibitem{b24}
K.~He, X.~Zhang, S.~Ren, and J.~Sun, ``Identity mappings in deep residual
  networks,'' in \emph{European conference on computer vision}.\hskip 1em plus
  0.5em minus 0.4em\relax Springer, 2016, pp. 630--645.

\bibitem{adam}
D.~P. Kingma and J.~Ba, ``Adam: A method for stochastic optimization,''
  \emph{arXiv preprint arXiv:1412.6980}, 2014.

\bibitem{Dublin}
A.~L. Goldberger, L.~A. Amaral, L.~Glass, J.~M. Hausdorff, P.~C. Ivanov, R.~G.
  Mark, J.~E. Mietus, G.~B. Moody, C.-K. Peng, and H.~E. Stanley, ``Physiobank,
  physiotoolkit, and physionet: components of a new research resource for
  complex physiologic signals,'' \emph{circulation}, vol. 101, no.~23, pp.
  e215--e220, 2000.

\bibitem{BestAIR2}
K.~Gleason, D.~Shin, M.~Rueschman, T.~Weinstock, R.~Wang, J.~Ware,
  M.~Mittleman, and S.~Redline, ``Challenges in recruitment to a randomized
  controlled study of cardiovascular disease reduction in sleep apnea: an
  analysis of alternative strategies. sleep,'' \emph{Sleep}, vol.~37, no.~12,
  pp. 2035--Äì2038, 2014.

\bibitem{BestAIR1}
G.~Zhang, L.~Cui, R.~Mueller, S.~Tao, M.~Kim, M.~Rueschman, S.~Mariani,
  D.~Mobley, and S.~Redline, ``The national sleep research resource: towards a
  sleep data commons,'' \emph{Journal of the American Medical Informatics
  Association}, vol.~25, no.~10, pp. 1351--Äì1358, 2018.

\bibitem{b22}
W.~Cai and D.~Hu, ``Qrs complex detection using novel deep learning neural
  networks,'' \emph{IEEE Access}, 2020.

\end{thebibliography}


\end{document}